\newcount\eqnumber
\def\chaphead{}

\def\new{\hbox{(\chaphead\the\eqnumber}\global\advance\eqnumber by 1}
\def\ref#1{\advance\eqnumber by -#1 (\chaphead\the\eqnumber
     \advance\eqnumber by #1 }
\def\first{\hbox{(\chaphead\the\eqnumber{a}}\global\advance\eqnumber by 1}
\def\last{\advance\eqnumber by -1 \hbox{(\chaphead\the\eqnumber}\advance
     \eqnumber by 1}
\def\eq#1{\advance\eqnumber by -#1 equation (\chaphead\the\eqnumber
     \advance\eqnumber by #1}
\def\eqnam#1{\xdef#1{\chaphead\the\eqnumber}}

\def\tcar{\futurelet\next\testnextcar}
\def\testnextcar{\ifhmode\ifcat\next.\else\ \fi\fi}

\def\gtorder{\mathrel{\raise.3ex\hbox{$>$}\mkern-14mu
             \lower0.6ex\hbox{$\sim$}}}
\def\ltorder{\mathrel{\raise.3ex\hbox{$<$}\mkern-14mu
             \lower0.6ex\hbox{$\sim$}}}

\catcode`\@=11 
\newif\iftwelv@  \twelv@true

\def\Textindent#1{\noindent\llap{#1\enspace}\ignorespaces}
\newdimen\referenceminspace  \referenceminspace=20pc
\def\testnextcar{\ifhmode\ifcat\next.\else\ \fi\fi}
\def\tcar{\futurelet\next\testnextcar}
\let\rel@x=\relax
\let\n@expand=\relax
\def\pr@tect{\let\n@expand=\noexpand}
\let\protect=\pr@tect
\let\gl@bal=\global
\newdimen\d@twidth
{\setbox0=\hbox{s.} \gl@bal\d@twidth=\wd0 \setbox0=\hbox{s}
        \gl@bal\advance\d@twidth by -\wd0 }
\def\removehglue{\loop \unskip \ifdim\lastskip >\z@ \repeat }
\def\roll@ver#1{\removehglue \nobreak \count255 =\spacefactor \dimen@=\z@
        \ifnum\count255 =3001 \dimen@=\d@twidth \fi
        \ifnum\count255 =1251 \dimen@=\d@twidth \fi
    \iftwelv@ \kern-\dimen@ \else \kern-0.83\dimen@ \fi
   #1\spacefactor=\count255 }
\def\step@ver#1{\rel@x \ifmmode #1\else \ifhmode
        \roll@ver{${}#1$}\else {\setbox0=\hbox{${}#1$}}\fi\fi }
\def\attach#1{\step@ver{\strut^{\mkern 2mu #1} }}
\def\spacecheck#1{\dimen@=\pagegoal\advance\dimen@ by -\pagetotal
   \ifdim\dimen@<#1 \ifdim\dimen@>0pt \vfil\break \fi\fi}
\newcount\referencecount     \referencecount=0
\newcount\lastrefsbegincount \lastrefsbegincount=0
\newif\ifreferenceopen       \newwrite\referencewrite
\newdimen\refindent          \refindent=30pt
\def\normalrefmark#1{\attach{\scriptstyle  #1  }}

\def\NPrefmark#1{\step@ver{{\;[#1]}}}
\def\refmark#1{\unskip\normalrefmark{#1}}
\def\refend@{\refmark{\number\referencecount}}
\def\refend{\refend@{}\space }
\def\refsend{\refmark{\count255=\referencecount
   \advance\count255 by-\lastrefsbegincount
   \ifcase\count255 \number\referencecount
   \or \number\lastrefsbegincount,\number\referencecount
   \else \number\lastrefsbegincount-\number\referencecount \fi}\space }
\def\REFNUM#1{\gl@bal\advance\referencecount by 1
    \xdef#1{\the\referencecount }}
\def\Refnum#1{\REFNUM #1\refend@ } 
\def\REF#1{\REFNUM #1\R@FWRITE\ignorespaces}
\def\Ref#1{\Refnum #1\REFWRITE }
\def\ref{\Ref\?}
\def\REFS#1{\REFNUM #1\gl@bal\lastrefsbegincount=\referencecount
    \REFWRITE }

\def\rf#1{\Ref#1}

\def\r@fitem#1{\par \hangafter=0 \hangindent=\refindent \Textindent{#1}}
\def\refitem#1{\r@fitem{#1.}}
\def\NPrefitem#1{\r@fitem{[#1]}}
\def\NPrefs{\let\refmark=\NPrefmark \let\refitem=NPrefitem}
\begingroup
 \catcode`\^^M=\active \let^^M=\relax %
 \gdef\p@rse@ndwrite#1#2{\begingroup \catcode`\^^M=12 \newlinechar=`\^^M%
         \chardef\rw@write=#1\sc@nlines#2}%
 \gdef\sc@nlines#1#2{\sc@n@line \g@rbage #2^^M\endsc@n \endgroup #1}%
 \gdef\sc@n@line#1^^M{\expandafter\toks@\expandafter{\deg@rbage #1}%
         \immediate\write\rw@write{\the\toks@}%
         \futurelet\n@xt \sc@ntest }%
\endgroup
\def\sc@ntest{\ifx\n@xt\endsc@n \let\n@xt=\rel@x
       \else \let\n@xt=\sc@n@notherline \fi \n@xt }
\def\sc@n@notherline{\sc@n@line \g@rbage }
\def\deg@rbage#1{}
\let\g@rbage=\relax    \let\endsc@n=\relax
\def\REFWRITE{\R@FWRITE\rel@x }
\def\R@FWRITE#1{\ifreferenceopen \else \gl@bal\referenceopentrue
     \immediate\openout\referencewrite=\jobname.refs
     \toks@={\begingroup \refoutspecials \catcode`\^^M=10 }%
     \immediate\write\referencewrite{\the\toks@}\fi
    \immediate\write\referencewrite{\noexpand\refitem %
                                    {\the\referencecount}}%
    \p@rse@ndwrite \referencewrite #1}
\def\refout{\bigskip\medskip
   \spacecheck\referenceminspace
   \ifreferenceopen \Closeout\referencewrite \referenceopenfalse \fi
   \hbox to\hsize{\hfil REFERENCES\hfil}\vskip.15in
   \input \jobname.refs
   }
\def\refoutspecials{\sfcode`\.=1000 \interlinepenalty=1000
         \rightskip=\z@ plus 1em minus \z@ }
\def\Closeout#1{\toks0={\par\endgroup}\immediate\write#1{\the\toks0}%
   \immediate\closeout#1}
\catcode`\@=12 
\documentstyle[12pt]{article}
\begin{document}
\def\etal{et al.}
\def\b8{\hbox{$^{8}$B}}
\tolerance=10000
\oddsidemargin -0.2363in
\evensidemargin -0.2363in
\topmargin -0.5625in
\headheight 12pt
\headsep 25pt
\footheight 12pt
\footskip 25pt
\textheight=9.5in
\textwidth=7in
\columnsep=.25in
\baselineskip=18pt
\ifnum\thepage=2 \textheight=5in\fi
\pagestyle{}
\twocolumn[\centerline{WHAT DO SOLAR MODELS TELL US}
\centerline{ABOUT SOLAR NEUTRINO EXPERIMENTS?}
\bigskip
\centerline{John N. Bahcall}
\centerline{Institute for Advanced Study}
\centerline{Olden Lane}
\centerline{Princeton, New Jersey}
\bigskip
\centerline{Abstract}
\bigskip
\centerline{\vbox{\hsize=5.5in
If the published event rates of the chlorine and Kamiokande
solar neutrino experiments are correct, then the energy spectrum of
neutrinos produced by the decay of $^8$B in the sun must be different
from the energy spectrum determined from laboratory nuclear physics
measurements.  This change in the energy spectrum requires physics
beyond the standard electroweak model.
In addition, the GALLEX and SAGE experiments, which
currently have large statistical
uncertainties, differ from the predictions of the standard solar model
by $2 \sigma$ and $3 \sigma$, respectively.}}\vglue-1.75in
]

At the conference, I presented a review of recent improvements in the
calculations of neutrino fluxes from solar models and then used the most
recent results to draw some conclusions about what we have learned by
comparing the results of  solar neutrino experiments with calculations
from solar models of the neutrino fluxes.
The analysis of solar models has now been published in detail
\rf\Pinsonneault{J. N. Bahcall, and M. H. Pinsonneault,
{\it Rev. Mod. Phys.}, {\bf 64}, 885, 1992.}, so there is no need to
repeat that material here.  My main goal at the conference was, in any
event, not to elucidate technical issues in solar model theory, but
rather to clarify and make quantitative the conclusions that folllow
from the confrontation of the solar model calculations with the four
operating experiments.
I will therefore take the model results as given and concentrate here on
what they teach us about the four solar neutrino experiments.

The first point to recognize is that the individual rates of the four
solar neutrino experiments tell us nothing about the possibility of new
physics until these rates are compared with solar models.
The analogy to an accelerator experiment is clear: we need to know what
the beam intensity is, as well as the flavor composition and energy
spectrum, in order to know if we are surprised or not by the
experimental rates.

The standard solar model
\rf\Bahcall{J. N. Bahcall and R. K. Ulrich,
{\it Rev. Mod. Phys.}, {\bf 60}, 297, 1988;
J. N. Bahcall, {\it Neutrino Astrophysics}
(Cambridge University Press, Cambridge, England, 1989).}
predicts the absolute fluxes from each of the important nuclear fusion
reactions and furthermore says that all solar neutrinos are
$\nu_e$'s.  What is more, to an accuracy of one part in $10^5$, the
energy spectrum of the $^8$B solar neutrinos must have the same shape as
the spectrum determined from laboratory nuclear physics experiments
\rf\Spectrum{J.N. Bahcall, {\it Phys. Rev. D.}, {\bf 44}, 1644, 1991.}.

The
invariance of the energy spectrum
allows us to compute
the rate of neutrino capture in the
chlorine experiment---independent of any considerations of
solar models---provided only that we know from the
Kamiokande experiment the flux of the
higher energy ($>~7.5$ MeV) $^8$B neutrinos.
In this process, we ignore the expected contributions to the chlorine
experiment, which has a threshold of only 0.8 MeV (an order of magnitude less
than the Kamiokande experiment), from $^7$Be, CNO, and pep neutrinos.
Using the empirical
result obtained for the Kamiokande experiment
\rf\Hirata{K. S. Hirata {\it et al.},
{\it Phys. Rev. Lett.}, {\bf 63}, 16, 1989; {\bf 65}, 1297, 1990.},
one finds that the predicted rate {\it from $^8$B neutrinos alone}
in the chlorine experiment is
6.20 SNU (from the standard model)$\times$0.48 (from the Kamiokande
measurement), or

\eqnam{\chlorineminrate}
$$
<\phi \sigma >_{\rm Cl; ~Kamiokande~only}  ~=~
$$
\vskip-24pt
$$
[3.0 \pm 0.3(1\sigma) \pm 0.4 ({\rm syst}) ~] ~ {\rm SNU}.
\eqno\new)
$$
This minimum rate, which ignores the contributions of all other neutrino
sources to the chlorine experiment, exceeds by $2\sigma$ the
observed chlorine rate,
\rf\Davis{
R. Davis Jr., in {\it Proc. of Seventh Workshop on Grand
Unification, ICOBAHN'86}, edited by J. Arafune (World
Scientific, Singapore, 1987),
p. 237; R. Davis Jr., K. Lande, C. K. Lee, P. Wildenhain, A.
Weinberger, T.  Daily, B. Cleveland, and J. Ullman, in {\it Proceedings of
the 21st International Cosmic Ray Conference}, Adelaide, Australia,
1990,
in press; J. K. Rowley, B. T. Cleveland, and R.
Davis Jr., in {\it Solar
Neutrinos and Neutrino Astronomy}, edited by  M. L. Cherry, W. A. Fowler, and
K. Lande (American Institute of Physics, New York, 1985), Conf. Proceeding No.
126, p. 1.}

\eqnam{\chlorinerate}
$$
<\phi \sigma >_{\rm Cl~exp}  ~=~ (2.2 \pm 0.2) ~ {\rm SNU},~~~
1 \sigma ~{\rm error} .
\eqno\new)
$$
Moreover, the lower-energy contributions from $^7$Be and $pep$ neutrinos
---which together amount to about 1.4 SNU---are much more reliably
determined by
 the theoretical calculations than
is the contribution from $^8$B neurinos.
If a fraction equal to 0.48 of the less-reliably calculated high
energy $^8$B neutrinos are
detected, then presumably more than 0.7 SNU of the expected 1.4 SNU from
pep and $^7$Be neutrinos should be added to the minimum rate of 3.0 SNU
calculated above.
On the basis of this comparison, Hans Bethe and I concluded
\rf\Betheone{J. N. Bahcall and H. A. Bethe, {\it Phys. Rev. Lett.},
{\bf 65}, 2233, 1990.}
that,
if the chlorine and Kamiokande experiments are both correct, then
physics beyond the standard electroweak model is required to change the
$^8$B neutrino energy spectrum.

More recently, Hans and I have sharpened this argument
\rf\Bethetwo{J. N. Bahcall and H. A. Bethe, {\it Phys. Rev. D.}, in press,
1992.} using a detailed Monte Carlo simulation of how the sun works.
The basis for our investigation is a collection of 1000
precise solar models \attach{\Bahcall} in which each input parameter
(the principal nuclear reaction
rates, the solar composition, the solar age, and the radiative opacity)
for each model was drawn randomly
from a normal distribution with the mean and standard deviation
appropriate to that variable.
The uncertainties in the neutrino cross sections \attach{\Bahcall}
for chlorine and for gallium were included by assuming a normal
distribution for each of the absorption cross sections with its estimated
mean and error.

We know that Monte Carlo simulations are necessary to understand the
results of complicated experiments in nuclear and particle physics.  It
should therefore seem natural to physicists that Monte Carlo simulations
are necessary to interpret the results of solar neutrino experiments;
the sun may be as complicated as a terrestrial
particle accelerator or detector.

The Monte Carlo study automatically
takes account of the nonlinear relations
among the different neutrino fluxes that are imposed by the coupled
partial differential equations of stellar structure
and by matching the stringent
boundary conditions of reproducing the observed solar luminosity, the heavy
element to hydrogen ratio, and the effective temperature at the present
solar age.
Attempts to simulate the uncertainties using average scaling laws
of the dependence of fluxes upon a single parameter, the central
temperature, can lead to serious errors.
A full Monte Carlo calculation is required to determine the
interrelations and absolute values
of the different solar neutrino fluxes.  For example, the fact that the
${\rm ^8B}$ flux may be crudely described as $\phi({\rm
^8B})~\propto~T^{18}_{\rm central}$ and $\phi({\rm
^7Be})~\propto~T^{8}_{\rm central}$ does not specify whether the two
fluxes increase and decrease together or whether their changes are out
of phase with each other.

Figure 1 shows the number of solar models with different predicted
event rates for the chlorine
solar neutrino experiment. The solar model with the best input
parameters predicts\attach{\Bahcall} an event rate of about 8 SNU.
None of the 1000 calculated solar models yields a capture
rate below 5.8 SNU.  Therefore, none of the 1000 solar models is within
$16 \sigma$ of the observed rate.
The discrepancy that is apparent in Figure~1 was for two decades the
entire ``solar neutrino problem.''
We can conclude from
Figure~1 that something is wrong
with either the standard solar model or the standard electroweak
description of the neutrino.

The largest and the most uncertain contribution to the predicted
chlorine rate is the $^8$B neutrino flux. This
quantity is completely unimportant for all astronomical purposes since
the reaction by which it is produced is extremely rare.  Suppose
therefore some mistake has been made in calculating the $^8$B neurino
flux and we normalize this flux, as before, by using the empirical
determination in the Kamiokande experiment.  What do we obtain for the
1000 solar models when we replace---for each model---the calculated flux
by a value determined
by the Kamiokande experiment?

Figure~2 provides a quantitative expression of the
difficulty in reconciling the Kamiokande and chlorine experiments by
changing solar physics, i.e., by arbitrarily changing the $^8$B neutrino
flux.
We constructed Figure~2 using the same 1000 solar models as were used in
constructing Figure~1, but for Figure~2 we artificially replaced the \b8 flux
for each standard model by a value drawn randomly
for that model from a normal distribution with
the mean and the standard deviation measured by Kamiokande.
The peak of the resulting distribution is moved
to 4.7 SNU (from 8 SNU) and the full width of the peak is
decreased by about a factor of three.
The peak is displaced because the measured (i.e., Kamiokande) value
of the \b8 flux is smaller than the calculated value.
The width of the distribution is decreased because
the error in the Kamiokande measurement is less than the
estimated theoretical uncertainty ($\approx 12.5\%$)
and because \b8 neutrinos constitute a smaller fraction of each displaced
rate than of the corresponding standard rate.

Figure~2 was constructed by assuming that something is seriously wrong
with the standard solar model, something that is
sufficient to cause the \b8 flux to be
reduced to the value measured in the Kamiokande experiment.
Nevertheless,
there is no overlap between the distribution of fudged standard
model rates and the measured chlorine rate.  None of
the 1000 fudged models lie within $3 \sigma$ (chlorine measurement
errors) of the
experimental result.

The results presented in Figures~1--2
suggest that new physics is required beyond the standard
electroweak theory if
the existing solar neutrino
experiments are correct within their quoted uncertainties.
Even if one abuses the solar models by artifically imposing
consistency with the
Kamiokande experiment, the resulting predictions of all 1000 of
the ``fudged'' solar models are inconsistent with the result of the
chlorine experiment(see Figure~2).

Figures~3a--3b show the number of solar models with different
predicted event rates for gallium detectors and
the recent measurements by the SAGE
\rf\Gavrin{V. N. Gavrin, \etal, {\it XXVI International
Conference on High Energy Physics}, Dallas, Texas, 1992;
A. I. Abazov {\it et al.},
{\it Phys. Rev. Lett.}, {\bf 67}, 332, 1991.}
 $(58^{+17}_{-24}~\pm~14 ({\rm syst})~{\rm SNU})$ and GALLEX
\rf\GALLEXone{P. Anselmann {\it et al.}, {\it Phys. Lett. B},
{\bf 285}, 376, 1992.}
$(83~\pm~19 (1 \sigma)~\pm~8 ({\rm syst})~{\rm SNU}$)
collaborations.
Figure 3a compares the gallium experimental results with the
``unfudged'' histogram of standard solar model calculations and Figure 3b
compares the results when the $^8$B neutrino flux is taken from the
Kamiokande mesurement.  Unlike the chlorine case (cf. Figures~1 and 2),
in which almost 80\%
of the predicted
event rate is from $^8$B neutrinos, Figures 3a and 3b are not
qualitatively different because $^8$B neutrinos  contribute very little
(only about 10\%)
to the predicted event rate in the gallium experiments.

With the current large statistical errors,
the gallium  measurements differ from the best-estimate theoretical
value
of 132~SNU
by approximately $2~\sigma$ (GALLEX) and $3.5~\sigma$
(SAGE). The gallium results provide modest support for the
existence of a solar neutrino problem, but by themselves do
not constitute a definitive conflict with standard theory.

\bigskip\medskip

\centerline{ACKNOWLEDGMENTS}
\medskip
This work was supported in part by the NSF via grant
PHY-91-06210 at I.A.S.

\refout

\end{document}